\documentclass[letter]{aa}
\usepackage{graphicx}
\usepackage[intlimits,sumlimits]{amsmath}
\usepackage{deluxetable}
\usepackage{times,epsfig} 
\usepackage{natbib}
\usepackage{amssymb}
\bibpunct{(}{)}{;}{a}{}{,} 
\usepackage{amsmath}
\usepackage[english]{babel}

\newcommand{\beqa}{\begin{eqnarray}} 
\newcommand{\eeqa}{\end{eqnarray}}

\newcommand{\bsub}{\begin{subequations}}
\newcommand{\esub}{\end{subequations}}
\newcommand{\beal}{\begin{align}}
\newcommand{\ealn}{\end{align}}
\newcommand{\Nif}{$\rm ^{56}Ni$} 
\newcommand{\Cif}{$\rm ^{56}Co$}
\newcommand{\Fif}{$\rm ^{56}Fe$}

\begin{document}
\title{Late-time emission of type~Ia supernovae: optical and 
near-infrared observations of SN~2001el 
\thanks{Based on observations collected at the European Southern Observatory,
Paranal, Chile (ESO Programmes 69.D-0193(ABCD) and 70.D-0023(AB).}}

\titlerunning{Late-time observations of SN~2001el.}
\authorrunning{Stritzinger \& Sollerman}
\author{M.~Stritzinger\inst{1}
 	\and J. Sollerman\inst{1,2}}

\institute{Dark Cosmology Centre, Niels Bohr
Institute, University of Copenhagen, Juliane Maries Vej 30, DK-2100 
Copenhagen \O, Denmark\\
\email{max,jesper @dark-cosmology.dk}
\and Stockholm Observatory, AlbaNova, Department of Astronomy, 106 91 
Stockholm, Sweden \\
} 

\offprints{M. Stritzinger}
\date{Received 21 December 2006 / Accepted May 6 2007}
\abstract{}{To elucidate the nature of the late-phase emission of the 
normal type Ia supernova SN~2001el.}
{We present optical and near-infrared light curves of SN~2001el 
from 310 to 445 days past maximum light, obtained 
with the Very Large Telescope.}
{The late-time optical ($UBVRI$) light curves decay in a nearly linear
fashion with decay time scales of 
$1.43\pm0.14$, $1.43\pm0.06$, $1.48\pm0.06$, $1.45\pm0.07$, and $1.03\pm0.07$
magnitudes (per hundred days) in the $U$, $B$, $V$, $R$ and $I$ bands, 
respectively. 
In contrast, in the near-infrared ($JHK{s}$) bands 
the time evolution of the flux appears to be nearly constant at these epochs. 
We measure decline rates (per hundred days) of 
$0.19\pm0.10$ and $0.17\pm0.11$ magnitudes in the 
$J$ and $H$ bands, respectively.
We construct a UVOIR light curve, and find that 
the late-time luminosity has a decay time scale nearly consistent with 
complete depostion of positron kinetic energy.}
{The late-time light curves of the normal type Ia SN~2001el demonstrate
the increased importance of the near-infrared contribution. This was previously observed
in the peculiar SN~2000cx, and the results for SN~2001el thus 
ensure that the conclusions previously based on a single peculiar 
event are applicable to normal type~Ia supernovae. 
The measured late-time UVOIR decline rate suggests that 
a majority of the positrons are trapped within the ejecta.
This results does not favor the predictions of a weak and/or radially 
combed magnetic field configuration.}

\maketitle
\section{Introduction}

Type Ia supernovae (hereafter SNe~Ia) are an 
exceptionally useful  tool for 
cosmological investigations 
\citep[see e.g.][ and references therein]{leibundgut01}. 
A large observational data set has been assembled, and major theoretical 
efforts have led towards an understanding of these stellar explosions.
However, most of the observational 
and theoretical focus has been directed towards the
understanding of SNe~Ia during the early photospheric phase. 
The study of the late-phases of thermonuclear supernovae, on the other
hand, offers other opportunities to study the physics of 
these explosions.

At times greater than $\sim$150 days past maximum light the ejecta are
transparent to the majority of $\gamma$ rays originating from the decay of the 
radioactive isotopes synthesized in the explosion. 
The light curves are then powered by the deposition of 
positron kinetic energy. 
The fraction of positron energy that is deposited into the ejecta is thought 
to depend on the magnetic field configuration; a strong and
tangled magnetic field should trap a majority of positrons whilst a weak 
and radially combed field would lead to a large positron escape fraction.

Late-time light curves therefore provide a way to understand the magnetic
field structure
\citep{colgate80,pilar98,milne99,milne01}.
If positrons are trapped (escape), a flattening
(steepening) in the bolometric light curve is expected. 
With this knowledge it should then be possible, at least in principle,
to constrain
the contribution these positrons make to the Galactic 511 KeV line.

The late-phases also allow for the study of other important aspects of the 
physics of SNe~Ia, such as the nucleosynthesis of 
radioactive \citep{stritzinger06} 
and stable \citep{kozma05} elements, as well as the distribution of these 
elements \citep[e.g.][]{motohara06}. In addition the late-phases allow us to
probe the physics of freeze-out and of the infrared catastrophe
\citep{axelrod80,fransson96,sollerman04}.

Early investigations proposed evidence for varying degrees of trapping 
\citep{cappellaro97,pilar98,milne99,milne01}. However, all such studies 
of the late-time emission were limited due to a lack of late-time
near-infrared data \citep[but see][]{elias83}.


A pioneering investigation of the optical and near-infrared 
late-time flux evolution of SN~2000cx 
found evidence for a less steep decrease of the bolometric light curve 
due to the increasing importance of the 
near-infrared contribution \citep{sollerman04}.
Although these findings for SN~2000cx were interesting for our understanding of positron trapping and the physics of the late-phases,  it was noted that SN\,2000cx was peculiar at early times \citep{li01,candia03}.
A concern was thus
that its late-phase behaviour may have been atypical compared to normal SNe~Ia. 
Here we conduct a similar study for a normal and more 
representative SN~Ia.

\subsection{SN~2001el}

SN~2001el was discovered on 17.0 September 2001 (UT) \citep{monard01} 
in the nearby spiral galaxy NGC~1448. NED gives a redshift to this
galaxy of 1168 km s$^{-1}$. The supernova was located $18\arcsec$ west 
and $20\arcsec$ north of the nucleus.
Four days after discovery, \citet{sollerman01} classified it as a bona fide 
SN~Ia.
This supernova has been targeted by several groups and a number of
papers are now available.
Early-phase light curves presented by \citet{krisciunas03} indicate
SN~2001el to be a normal event with a $\Delta$m$_{15}$($B$) $=$ $1.13\pm0.04$ mag. 
The time of $B$-band maximum was JD 2,452,182.5 or
30.0 September 2001.
\citet{sollerman05} estimated the host galaxy extinction to be
E($B-V$)$_{\rm host} = 0.18\pm0.08$ mag, and the dust maps of 
\citet{schlegel98} list the Galactic component to be E($B-V$)$_{\rm gal} = 
0.014$ mag.   
\citet{mattila05} published early- and late-time   
spectroscopy obtained with the Very Large Telescope (VLT). 

We selected SN~2001el for a VLT study since it was nearby, 
well-studied at early phases, and 
located in a favorable position within the host galaxy. The sky position
 allowed a
dedicated multi-epoch multi-band observational campaign to be conducted 
at the VLT.
In this letter we present these late-time optical and near-infrared 
photometric observations  of SN~2001el.


\section{Observations and data analysis}
\subsection{Optical imaging with the VLT}

Late-time multi-epoch $UBVRI$-band observations of SN~2001el were obtained
with the FOcal Reducer and low dispersion Spectrograph (FORS1) attached to
UT1 (Aug.--Oct. 2002) and UT3 (Nov.--Dec. 2002) of the VLT.
Service mode imaging was conducted from 310 to 437 days past 
$B$-band maximum. Table 1 contains a log of the optical observations.

FORS1 uses a $2048$ $\times$ $2048$ pixel sized charged 
coupled device with a pixel scale of $0\farcs20$ per pixel. 
To obtain an adequate signal-to-noise while preventing saturation of field 
stars, multiple exposures were obtained in each filter. 
All frames were bias-subtracted and twilight-sky flattened using 
{\tt MIDAS}. 
Frames taken for a given night with the same filter were then combined 
with IRAF\footnote{The Image Reduction and Analysis
Facility (IRAF) is maintained and distributed by the Association of 
Universities for Research in Astronomy, under a cooperative agreement with the
National Science Foundation.} scripts. 

The brightness of SN~2001el was determined differentially with respect to 
a sequence of local field stars. 
Absolute photometry of the local sequence was obtained using 
standard stars in the fields Mark~A and PG41528+062 \citep{landolt92}, which were 
observed under photometric conditions on 5.2 August  2002. 
Instrumental magnitudes of the standard fields 
were computed using the IRAF {\tt DAOPHOT} package {\tt phot} with 
an aperture radius of 7$\arcsec$. The instrumental magnitudes 
were then used to derive the zero-points used to put the local sequence on the
standard system. 

Instrumental magnitudes of 10 stars in the local field 
were computed using {\tt phot} with an aperture radius of 0$\farcs$5.
Aperture corrections were
then computed using the IRAF task {\tt mkapfile}. 
The instrumental magnitudes were 
color term and atmospheric absorption corrected and then put on the
standard system using the zero-points computed from the Landolt fields.
Table~2 lists their standard magnitudes, where the listed uncertainties 
include errors associated with the photometric zero-point and the errors
computed by {\tt phot} and  {\tt mkapfile}.
Note that in the $U$ band only 8 of the sequence stars are bright enough 
to calculate an accurate magnitude. 
These local sequence stars showed a night-to-night dispersion of less
than 0.05 mag (std. dev.) over the entire duration of the programme.
 
Instrumental magnitudes of the supernova were measured using  
{\tt phot} with an aperture radius of $0\farcs5$. 
When using the Landolt standards to solve for the photometric solutions we 
found that the color terms and extinction coefficients were
very similar to the averaged values provided in the ESO
web-page.\footnote{http://www.eso.org/observing/dfo/quality/index\_fors1.html} 
As the ESO values are based on a significantly larger number
of standard stars we decided to adopt these averaged values. This allowed us
to reduce the number of fitting parameters such that it was only necessary to
compute the zero-point for each epoch using the local sequence. 

Table~5 contains the $UBVRI$-band photometry of SN~2001el. 
The quoted uncertainties were calculated by adding in quadrature the errors 
associated with the nightly zero-point, and the errors computed by {\tt phot} and
{\tt mkapfile}. 

\subsection{Infrared observations with the VLT}

Late-time multi-epoch $JHK{s}$-band observations were obtained with the 
Infrared Spectrometer And Array Camera (ISAAC)\footnote{http://www.eso.org/
instruments/isaac/} attached to UT1 of the VLT. 
ISAAC is equipped with a
1024$\times$1024 pixel sized Rockwell Hawaii HgCdTe detector with
a pixel scale of 0$\farcs$147 per pixel and a 
2$\farcm$5$\times$2$\farcm$5 field of view. 
Imaging was conducted in the short wavelength mode from 316 to 445 days
past maximum light. A log of the observations is given in Table~3.

As SN~2001el was fairly close to the galaxy, imaging was
performed using the jitter--offset method. The procedure consists of obtaining
typically two images of the galaxy with a small offset of the telescope 
between each frame. The telescope is then dithered off source
to obtain an image of the sky. Multiple iterations of jitter--offset
are performed and a data cube for each filter is obtained.

All data were reduced using the {\tt Eclipse}\footnote{http://www.eso.org/
projects/aot/eclipse/} software package. 
The task {\tt jitter} was used to estimate and remove the sky background, 
and then combine each on-source image within the data cube. 

A local sequence consisting of 7 stars in the field around NGC~1448 was
calibrated with standard fields. 
Instrumental magnitudes of the standard stars were obtained using
{\tt phot} with an aperture radius of 0$\farcs$5 and aperture corrections were
then computed with {\tt mkapfile}.
In the $J$ band the local sequence
was calibrated using the standard fields FS31 and FS32 
\citep{hawarden01,leggett06},
which were observed on 16.5 December 2002, while observations of the 
standard star field FS19 obtained on 15.8 December 2002 was used
to calibrate the local sequence in the $H$ band. 
Three of our local sequence stars were observed by 
the 2MASS all-sky survey. We found that our calibrated $JH$-band magnitudes
of the 2 brightest of these 3 stars agreed with the 2MASS value to within
0.1 mag. For the $K_{s}$ band we were unable to obtain a zero-point from 
the observed standard fields that 
gave magnitudes consistent with the 2MASS observations. 
We therefore calibrated the local sequence in the $K_{s}$ band using the 
two bright sequence stars that are in common with the 2MASS catalog. 

Magnitudes of the photometric sequence (corrected for atmospheric 
extinction\footnote{The photometry was corrected for atmospheric absorption
using standard extinction coefficients: http://www.eso.org/instruments/isaac/imaging\_standards.html})
are listed in Table~4. No color term corrections were applied.
The quoted uncertainties are added in quadrature 
using the errors in the zero-points, and the errors computed by {\tt phot} and {\tt mkapfile}. 
We found the night-to-night dispersion of each sequence star over the course
of the programme to be less than 0.05 mag (std. dev.).

Aperture photometry of the supernova was computed
with {\tt phot} using an aperture radius of 0$\farcs$5  and then
aperture corrected. 
Table~6 lists the $JHK_{s}$-band photometry. The quoted
uncertainties were determined by adding in quadrature the
errors associated with the nightly zero-point,
the photometric error computed by {\tt phot}, the error computed 
by {\tt mkapfile}, and a 0.10 magnitude error 
(for $JH$ bands) associated with the 
difference between our standard magnitudes and the 2MASS magnitudes.

\section{Results}


In Fig.~1 we present optical and near-infrared light curves of
SN~2001el. Here the early-time data from \citet{krisciunas03} are plotted
together with our late-time ($>$300 days past $B_{\rm max}$) VLT photometry.
No extinction corrections have been applied to these light curves. 
As seen in Fig.~1 the late-time optical light curves follow a nearly
linear decline, while the near-infrared light 
curves are essentially flat. We measure decline rates (mag per hundred days)
of $\simeq$1.45 for the  $UBVR$ light curves, a shallower 1.03 in the 
$I$ band, $\simeq$0.20 in the $J$ and $H$ bands, and $\simeq$$-$1.0$\pm0.7$
in the $K_{s}$ band.
Table 7 lists the exact late-time decline rates for all passbands. 
A final $K_{s}$-band image was obtained 443 days 
past maximum light. However, the final sky-subtracted image was not
deep enough to provide any intensity limit on the supernova 
brightness at this epoch.

\begin{figure}
\resizebox{\hsize}{!}{\includegraphics{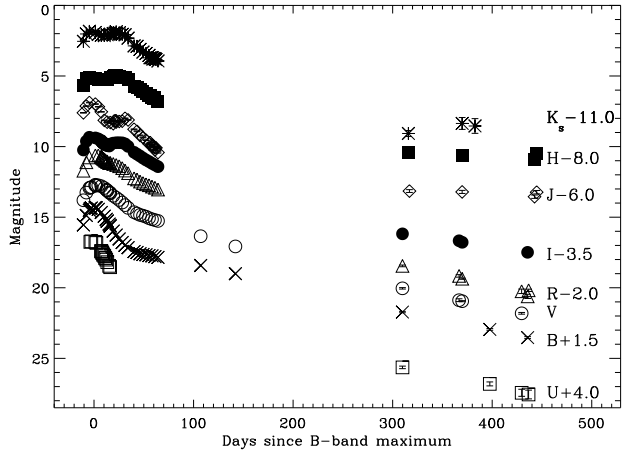}}
\caption{$UBVRIJHK_{s}$ light curves of SN~2001el plotted as a function of time
since $B$-band maximum. Early-time photometry ($<$200 days) is
from \citet{krisciunas03}.
The data have been shifted in the y-direction for clarity as indicated.
}
\end{figure}

We constructed an UltraViolet Optical near-InfraRed (UVOIR) bolometric light curve following the
method of \citet{contardo00}; see Fig.~2. 
To place the  UVOIR flux on an absolute flux scale we adopted
the extinction given in Sect. 1.1 and a distance to NGC~1448 of 17.9 Mpc 
\citep{mattila05}. 

With the late-phase UVOIR light curve we can estimate the 
fraction of flux (as a function of time) emitted in the $JHK_{s}$ 
passbands. It is found that the percentage of UVOIR flux in these passbands
during 310 to 445 days past maximum light increases from $\sim$6\% to 
$\sim$25\%.

A least-squares fit to the late-time UVOIR light curve 
yields a decline
rate (per hundred days) of 0.90 mag. This is consistent with 
the 0.98 mag decline rate expected for the case of complete deposition of 
positron kinetic energy.
\begin{figure}
\resizebox{\hsize}{!}{\includegraphics{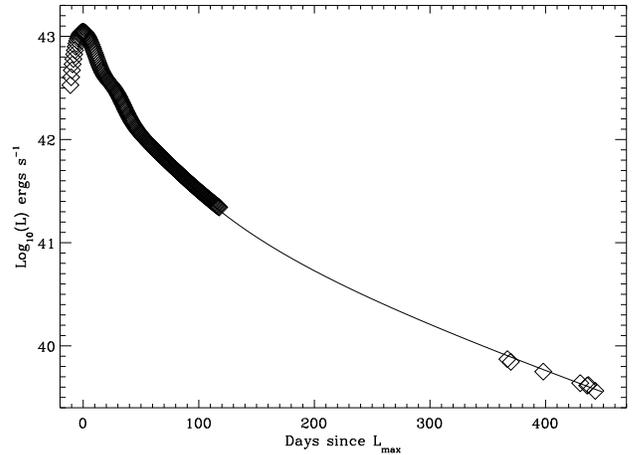}}
\caption{The UVOIR light curve (diamonds) fitted with our toy model
(full line).}
\end{figure}

Finally, to estimate the \Nif~mass we performed a least-squares 
fit of a radioactive decay energy deposition 
function for the \Cif$\rightarrow$\Fif~decay chain to our
UVOIR light curve between 50 and 450 days past maximum brightness.
The functional form of the toy model used \citep[see][]{sollerman98} is 
$L~=~1.3 \times 10^{43}\, M_{\rm Ni}~ 
\mathrm{e}^{-t/111.3}  \left(1 - 0.966\mathrm{e}^{-\tau}\right)$,
in units of erg s$^{-1}$, where the optical depth, $\tau$,
is given by (t$_{1}$/t)$^{2}$. In this case t$_{1}$ sets the time
when the optical depth to $\gamma$ rays is unity. We find that a best fit 
to the data is obtained with a t$_{1}$ of 35 days and a \Nif~mass of 
0.40 M$_{\sun}$ (see Fig. 2). 

By comparing this \Nif~mass to the estimate presented   
by \citet{stritzinger06} (corrected to the same absolute flux scale used 
in this work), which used Arnett's rule,
we can constrain the amount of flux not encapsulated in our late-time
UVOIR light curve.
We find that $\sim$40\% of the total flux is emitted
outside the covered passbands. This value is consistent with 
results obtained from both observations and modeling of 
SN~2000cx \citep{sollerman04}. 
It must be noted however that our toy model, Arnett's rule, and the 
modeling of SN~2000cx are all based on a number of uncertain assumptions,
so these results should only be considered as indicative. It is important
to keep in mind that we are still not probing the true bolometric light
curve. 

\section{Discussion} 

This is the first normal SN~Ia for which we have systematically monitored the
late-time light curves in both the optical and the near-infrared.
The optical late-time behaviour of SN~2001el 
is nearly identical to that of 
other SNe~Ia that have been studied at late-phases.
The $BVRI$-band decline rates are in agreement with the average
decline rates presented by \citet{milne01} and \citet{lair06} for a 
number of normal SNe~Ia. Our $U$-band light curve 
shows a decline rate similar to the other optical bands
\citep[but see][]{stanishev07}.

The shallow decline in the $I$ band suggests a shift of flux towards the
near-infrared, as also indicated by \citet{sollerman04} and \citet{lair06}.
This is confirmed by our late $JHK_{s}$ light curves. 
After $\sim$300 days past maximum light the emission redward of 
1.1 microns becomes increasingly important. 
This behaviour was documented with the detailed observations of
SN~2000cx \citep{sollerman04}. 
Our observations of SN~2001el indicate that
at late-phases SN~2000cx behaved like a {\it normal} SN~Ia.
Similar behaviour was seen with 
limited near-infrared observations of three other normal SNe~Ia: SN~1998bu 
\citep{spyromilio04}, SN~2003cg \citep{eliasrosa06}, and SN~2004S \citep{krisciunas07}. 

The 
decline rate of the late-time UVOIR light curve indicates that 
few positrons escape the ejecta. 
This result does not favor predictions of a weak and/or radially
combed magnetic field configuration.

We have presented modeling of the late-phase observations of SN~2000cx and
we refer the reader to  \citet{sollerman04} for a general discussion
concerning the physics driving these observations. 
The main result of the current paper 
is the demonstration that the increasing importance of the 
late-time near-infrared light curves is indeed a generic feature of 
normal SNe~Ia. This result challenges the assumption used by 
other studies that
the late-time bolometric light curve follows the optical light curves.
Therefore any hitherto conclusions regarding positron escape and their
contribution 
to the Galatic 511~KeV line based on optical 
photometry  alone must be reevaluated. These data will serve as
sorely needed input for modeling that will attempt to address this 
issue.

\begin{acknowledgements}
The Dark Cosmology Centre is funded by the 
Danish National Research Foundation. 
This research was supported in part by the National Science Foundation under
Grant No. NSF PHY05-51164.
Special thanks to the referee for a prompt and useful report.
We thank Johan Fynbo and Palle Jakobsson for helpful discussions on data
reductions,
and also Bruno Leibundgut, Peter Lundqvist, Kevin Krisciunas, Peter Milne,
and Dong Xu for discussions related to this project.
This research has made use of the Two Micron All-Sky
Survey, and the NASA/IPAC 
Extragalactic Database (NED), which is operated by the Jet Propulsion 
Laboratory, California Institute of Technology, under contract with the 
National Aeronautics and Space Administration.

\end{acknowledgements}


\clearpage 
\setcounter{table}{0}
\begin{table}
\caption{Log of optical VLT observations for SN~2001el.}
\label{table:1}
\centering
\begin{tabular}{ccccc}
\hline\hline
Phase$^{a}$ & Filter & Exposure & Airmass & Seeing  \\
(days)      &        & (s)      &         & (arcsec) \\
\hline
310        & $U$  & 2$\times$800 & 1.24 & 0.89 \\
310        & $B$  & 3$\times$180 & 1.41 & 0.56  \\
310        & $V$  & 3$\times$150 & 1.32 & 0.62  \\
310        & $R$  & 3$\times$150 & 1.28 & 0.66  \\
310        & $I$  & 3$\times$180 & 1.25 & 0.56  \\
367        & $V$  & 3$\times$300 & 1.23 & 1.16  \\
367        & $R$  & 3$\times$300 & 1.19 & 0.98  \\
367        & $I$  & 3$\times$300 & 1.15 & 1.08  \\
370        & $V$  & 3$\times$300 & 1.08 & 0.60  \\
370        & $R$  & 3$\times$300 & 1.19 & 0.60  \\
370        & $I$  & 3$\times$300 & 1.11 & 0.60  \\
398        & $U$  &2$\times$1020 & 1.36 & 1.00 \\
398        & $B$  & 3$\times$300 & 1.26 & 0.78 \\
430        & $U$  & 1$\times$790 & 1.07 & 1.02 \\ 
430        & $V$  & 3$\times$600 & 1.11 & 0.80 \\
430        & $R$  & 3$\times$720 & 1.25 & 1.00 \\
436        & $B$  & 3$\times$600 & 1.07 & 0.84 \\
436        & $R$  & 3$\times$720 & 1.07 & 1.20 \\
436        & $I$  & 6$\times$900 & 1.19 & 1.00 \\
437        & $U$  &3$\times$1020 & 1.20 & 1.00\\ 
437        & $R$  & 3$\times$720 & 1.11 & 0.80 \\
\hline
\end{tabular} \\
\begin{tabular}{lll}
$^a$ Refers to days past $B_{\rm max}$. && \\
\end{tabular}
\end{table}

\setcounter{table}{1}
\begin{table}
\caption{Magnitudes for local standards in the optical.}
\label{table:2}
\centering
\begin{tabular}{rrccccc}
\hline\hline
Offsets$^a$ & & $U$ & $B$ & $V$ & $R$ & $I$ \\
\hline
181.6 N & 72.8 E  & 20.50(0.08)$^b$  & 20.68(0.05) & 20.13(0.05) & 19.78(0.06) & 19.43(0.05) \\
193.6 N & 61.6 E  & 19.84(0.08)      & 19.84(0.05) & 19.18(0.05) & 18.77(0.06) & 18.40(0.05) \\
179.2 N & 56.8 W  & 20.66(0.08)      & 20.66(0.05) & 20.00(0.05) & 19.59(0.06) & 19.21(0.05) \\
6.4   N & 113.6 W & 20.42(0.08)      & 20.33(0.05) & 19.61(0.05) & 19.19(0.06) & 18.80(0.05) \\
56.8  S & 99.2 W  & 22.44(0.09)      & 21.41(0.05) & 19.73(0.05) & 18.70(0.06) & 17.60(0.05) \\
62.4 S  & 95.2 W    & 22.73(0.10)      & 22.69(0.05) & 22.01(0.05) & 21.66(0.06) & 21.32(0.06) \\
72.8 S  & 140.0 W & 19.51(0.08)      & 19.67(0.05) & 19.18(0.05) & 18.86(0.06) & 18.54(0.05) \\
73.6 N  & 181.6 E & 20.73(0.08)      & 21.10(0.05) & 21.02(0.05) & 20.92(0.06) & 20.71(0.05) \\
39.2 S  & 136.8 E & \nodata          & 21.57(0.05) & 20.09(0.05) & 19.19(0.06) & 18.33(0.05) \\
132.0 S & 22.4 W  & \nodata          & 23.18(0.05) & 21.62(0.05) & 20.67(0.06) & 19.68(0.05) \\
\hline
\end{tabular} \\
\begin{tabular}{lll}
$^a$ Offsets in arcseconds measured from the supernova. && \\
$^b$ Numbers in parentheses are uncertainties. && \\
\end{tabular}
\end{table}

\setcounter{table}{2}
\begin{table}
\caption{Log of near-infrared VLT observations for SN~2001el.}
\label{table:3}
\centering
\begin{tabular}{ccccc}
\hline\hline
Phase$^{a}$ & Filter & Exposure$^b$ & Airmass & Seeing  \\
(days)      &        & (s)      &         & (arcsec) \\
\hline
316        & $H$     &   10$\times$6$\times$25 & 1.13 & 0.49 \\
316  & $K_{s}$ &   10$\times$6$\times$28 & 1.30 & 0.41 \\
317        & $J$     &   30$\times$4$\times$24 & 1.18 & 0.52 \\
370        & $J$     &   30$\times$4$\times$23 & 1.06 & 0.65 \\
370        & $H$     &   10$\times$6$\times$17 & 1.15 & 0.44 \\
370        & $K_{s}$ &   10$\times$6$\times$30 & 1.08 & 0.40 \\
383        & $K_{s}$ &   10$\times$6$\times$20 & 1.11 & 0.40 \\
443        & $H$     &   10$\times$6$\times$30 & 1.14 & 0.40  \\
443        & $J$ &   30$\times$4$\times$30 & 1.14 & 0.40 \\
443  & $K_{s}$ &   30$\times$4$\times$27 & 1.09  & 0.60  \\
445        & $J$ &   30$\times$4$\times$23 & 1.17 & 0.70 \\
445        & $H$     &   10$\times$6$\times$60 & 1.06 & 0.55 \\
\hline
\end{tabular} \\
\begin{tabular}{lll}
$^a$ Refers to days past $B_{\rm max}$. && \\
$^b$ Detector integration time (DIT)$\times$number of DITs per exposure$\times$
number of exposures. && \\
\end{tabular}
\end{table}
\clearpage

\setcounter{table}{3}
\begin{table}
\caption{Magnitudes for local standards in the near-infrared.}
\centering
\begin{tabular}{llccc}
\hline\hline
Offsets$^a$ &  & $J$ &  $H$ & $K_{s}$ \\
\hline
55.35 N & 11.96 W     &   12.98(0.06)$^{b}$ & 12.57(0.05)  &  12.51(0.05)\\
     41.9 N   &      21.4 W &         17.54(0.06)  &      16.89(0.05)   &  16.09(0.05)\\
17.4 S  & 46.3 W & 13.56(0.06)      & 13.18(0.05) & 13.07(0.05)\\
25.4 S  & 42.3 E & 17.54(0.06)      & 17.24(0.05) & 16.95(0.05)\\
53.9 S  & 24.5 E & 19.36(0.06)      & 18.91(0.06) & 18.75(0.05)\\
50.8 S  & 62.8 E & 16.83(0.06)      & 16.44(0.05) & 16.24(0.05)\\
65.9 S  & 49.4 E & 16.27(0.06)      & 15.83(0.05) & 15.59(0.05)\\
\hline
\end{tabular} \\
\begin{tabular}{lll}
$^a$ Offsets in arcseconds measured from the supernova. && \\
$^b$ Numbers in parentheses are uncertainties. && \\
\end{tabular}
\end{table}

\setcounter{table}{4}
\begin{table}
\caption{Late-time optical magnitudes of SN~2001el.}
\centering
\begin{tabular}{cccccc}
\hline\hline
Phase$^{a}$ & $U$ & $B$  & $V$ & $R$ & $I$ \\
(days)      &     &      &     &     &     \\
\hline
 310 &21.64(0.09)$^b$  & 20.22(0.05) &  20.04(0.05)  & 20.45(0.06) & 19.68(0.05) \\
 367 & \nodata      & \nodata      &  20.88(0.10)  & 21.16(0.09) & 20.17(0.10) \\
 370 & \nodata      & \nodata      &  20.95(0.04)  & 21.36(0.04) & 20.28(0.03) \\
 398 &22.81(0.14)  & 21.45(0.06) & \nodata       & \nodata      & \nodata      \\
 430 &23.44(0.25)  & \nodata      &  21.81(0.06)  & 22.25(0.16) & \nodata      \\
 436 & \nodata      & 22.03(0.07) & \nodata       & 22.61(0.15) & 20.99(0.08) \\
 437 &23.56(0.31)  & \nodata      & \nodata       & 22.18(0.08) & \nodata      \\
\hline
\end{tabular} \\
\begin{tabular}{lll}
$^a$ Refers to days past $B_{\rm max}$. && \\
$^b$ Numbers in parentheses are uncertainties. && \\
\end{tabular}
\end{table}

\setcounter{table}{5}
\begin{table}
\caption{Late-time near-infrared magnitudes of SN~2001el.}
\centering
\begin{tabular}{cccc}
\hline\hline
Phase$^a$ & $J$ & $H$ & $K_{s}$ \\
(days)    &     &     &        \\
\hline
 316 & \nodata          & 18.40(0.12)$^{b}$ &  20.07(0.21)  \\
 317 & 19.15(0.10)      &  \nodata          &  \nodata      \\
 370 & 19.21(0.11)      & 18.62(0.11)       &   19.36(0.42)\\
 383 & \nodata       &  \nodata          &   19.54(0.52) \\
 443 & 19.55(0.12)      & 18.89(0.11)       &   \nodata      \\
 445 & 19.23(0.12)      & 18. 47(0.12)       &   \nodata       \\
\hline
\end{tabular} \\
\begin{tabular}{lll}
$^a$ Refers to days past $B_{\rm max}$. && \\
$^b$ Numbers in parentheses are uncertainties. && \\
\end{tabular}
\end{table}

\setcounter{table}{6}
\begin{table}
\caption{Decline rates of late-time light curves$^a$.}
\label{table:7}
\centering
\begin{tabular}{cccccccc}
\hline\hline
$U$ & $B$ & $V$ & $R$ & $I$ & $J$ & $H$ & $K_{s}$ \\
\hline
1.43(0.14) &  1.43(0.06) &  1.48(0.06) &  1.45(0.07) &  1.03(0.07) &  0.19(0.10) &  0.17(0.11) &  -1.04(0.65)\\
\hline
\end{tabular}
\begin{tabular}{lll}
$^a$ Mag per 100 days between 310 and 450 days; errors in parenthesis are $1\sigma$. && \\
\end{tabular}
\end{table}

\end{document}